\documentclass[12pt]{article}
\pdfoutput=1
\usepackage{jheppub}
\usepackage[utf8]{inputenc}
\usepackage{verbatim}
\usepackage{amsmath}
\usepackage{amssymb}
\usepackage{amsthm}
\usepackage{slashed}
\usepackage{amsfonts}
\usepackage{color}
\usepackage{xcolor}

\newcommand{\be}{\begin{equation}}
\newcommand{\ee}{\end{equation}}
\newcommand{\bfig}{\begin{figure}\begin{center}}
\newcommand{\efig}{\end{center}\end{figure}}
\newcommand{\bi}{\begin{itemize}}
\newcommand{\ei}{\end{itemize}}

\newcommand{\Tr}{\mathrm{Tr}}

\newcommand{\ol}{\overline}

\theoremstyle{definition}

\begin{document}
\preprint{MIT--CTP 5249}
\title{Global symmetry, Euclidean gravity, and the black hole information problem}
\author[a]{Daniel Harlow}
\author[b]{and Edgar Shaghoulian}
\affiliation[a]{Center for Theoretical Physics\\ Massachusetts Institute of Technology, Cambridge, MA 02139, USA}
\affiliation[b]{David Rittenhouse Laboratory, University of Pennsylvania\\
209 S.33rd Street, Philadelphia PA, 19104, USA}
\emailAdd{harlow@mit.edu, eshag@sas.upenn.edu}
\abstract{In this paper we argue for a close connection between the non-existence of global symmetries in quantum gravity and a unitary resolution of the black hole information problem.  In particular we show how the essential ingredients of recent calculations of the Page curve of an evaporating black hole can be used to generalize a recent argument against global symmetries beyond the AdS/CFT correspondence to more realistic theories of quantum gravity.  We also give several low-dimensional examples of quantum gravity theories which do not have a unitary resolution of the black hole information problem in the usual sense, and which therefore can and do have global symmetries.  Motivated by this discussion, we conjecture that in a certain sense Euclidean quantum gravity is equivalent to holography.\\ \begin{center}{\small \emph{Dedicated to Leonard Susskind on the occasion of his 80th birthday}}\end{center}}
\maketitle

\section{Introduction}
The idea that in quantum gravity there should be no global symmetry has a long history \cite{Banks:1988yz,Giddings:1988cx,Kallosh:1995hi,ArkaniHamed:2006dz,Banks:2010zn,Harlow:2018jwu,Harlow:2018tng}.  The most naive argument for this is simply that global charge can be thrown into a black hole, after which there is no record of it outside.  This argument however neglects the possibility that the charge is stored in the black hole, eventually coming out either in the radiation or being left in some evaporation remnant.  In the case of a continuous global symmetry this loophole can be removed provided we assume the validity of the Bekenstein-Hawking formula
\be\label{BHform}
S_{BH}=\frac{\mathrm{Area}}{4G}
\ee
for the coarse-grained entropy of a black hole (or more generally the Wald formula \cite{Wald:1993nt}), since a black hole with finite number of microstates cannot store an arbitrarily large amount of information about its initial global charge \cite{Banks:2010zn}.  This argument also applies to infinite discrete  groups such as $SL(2,\mathbb{Z})$, but it does not apply to finite groups such as $\mathbb{Z}_2$ because for such groups the number of irreducible representations is finite.  More recently, in \cite{Harlow:2018jwu,Harlow:2018tng} an argument against all global symmetries, both discrete and continuous, was given, but only within the context of the AdS/CFT correspondence.  

The general trend of these arguments is that the conclusions have been getting stronger, but so have the assumptions. Since we do not live in a universe with negative cosmological constant, we would like to replace the assumption of AdS/CFT with something weaker.  Whatever assumption we adopt must be nontrivial: for a sufficiently permissive definition of ``quantum gravity'', there are quantum gravity theories which \textit{do} have global symmetries! So far all known examples live in fewer than $3+1$ spacetime dimensions, and there is currently no reason to expect any $3+1$ (or higher) dimensional examples with a propagating graviton to exist, but nonetheless we clearly need to impose some kind of requirement to exclude these theories from any argument against global symmetry.

The proposal of this paper is that the key property a theory of quantum gravity must have to forbid the existence of global symmetries is unitary black hole evaporation which is consistent with the Bekenstein-Hawking entropy formula \eqref{BHform}. We will not rigorously establish this statement, but we will provide evidence for it from a variety of directions.  To begin with, the validity of \eqref{BHform} is already sufficient to exclude continuous global symmetries by the argument of \cite{Banks:2010zn}.\footnote{See the introduction of \cite{Harlow:2018tng} for a more detailed review of that argument.  The key point in making sure that a black hole of finite mass can be prepared with arbitrarily large charge is that the Hawking process produces no net flux of global charge away from an evaporating black hole, and even in a theory where the evaporation is unitary we expect that Hawking's framework correctly computes low-point correlation functions such as the flux of global charge.} Moreover black hole evaporation is unitary and consistent with \eqref{BHform} in the AdS/CFT correspondence \cite{Witten:1998qj,Maldacena:2001kr,Harlow:2014yka}, so our proposal is consistent with the arguments of \cite{Harlow:2018jwu,Harlow:2018tng}.  Finally the low-dimensional theories of quantum gravity with global symmetries that we present below do \textit{not} have black holes obeying \eqref{BHform}, which is again consistent with our proposal.  We expect that \eqref{BHform} and unitarity do hold in any sufficiently semiclassical compactification of string theory, so our proposal is also consistent with the fact that no string vacuum with a global symmetry has ever been discovered (continuous global symmetries are not possible in perturbative string theory \cite{Banks:1988yz}, but in the discrete case this is more of an ``empirical'' observation).

In addition to the circumstantial evidence described in the previous paragraph, we can also give a direct argument that, under a set of assumptions we state below, unitary black hole evaporation consistent with the Bekenstein-Hawking formula implies the non-existence of global symmetries. Our argument makes use of the recent discovery that the ``quantum extremal surface'' (QES) formula of \cite{Engelhardt:2014gca} is powerful enough to compute a unitary Page curve for an evaporating black hole in a wide variety of circumstances \cite{Almheiri:2019psf,Penington:2019npb,Almheiri:2019yqk,Rozali:2019day,Almheiri:2019psy,Bousso:2019ykv,Gautason:2020tmk,Anegawa:2020ezn,Hartman:2020swn,Balasubramanian:2020hfs,Geng:2020qvw,Hollowood:2020cou,Krishnan:2020fer,Bak:2020enw,Chen:2020jvn,Chen:2020hmv,Chen:2020tes,Hartman:2020khs,Balasubramanian:2020xqf}. In this paper we will assume that the basic logic of these calculations applies in any situation where a black hole which is formed quickly evaporates in a unitary manner. We will argue that in any theory where this happens, no global symmetries are possible.  Our argument is essentially a recasting of the argument in \cite{Harlow:2018jwu,Harlow:2018tng}, replacing pieces of the boundary CFT with pieces of the Hawking radiation (this replacement can be thought of as running the analogy between these pieces advocated in \cite{Akers:2019nfi} in reverse).

Currently the strongest arguments for the validity of the QES formula in these Page curve calculations are based on the Euclidean gravity path integral \cite{Almheiri:2019qdq,Penington:2019kki}, building on \cite{Lewkowycz:2013nqa,Faulkner:2013ana,Dong:2016hjy,Dong:2017xht}.  This is also the most general method for justifying the Bekenstein-Hawking formula \eqref{BHform} \cite{Gibbons:1976ue}, and our expectation is that these results go hand-in-hand.  We will see a close relationship between Euclidean quantum gravity and the information problem at several points in this note, and in our final discussion we will interpret this in terms of a conjectured equivalence between holography and Euclidean quantum gravity.

The structure of this paper is as follows. In section \ref{lowdimsec} we review a few low-dimensional examples of quantum gravity theories with global symmetries, emphasizing that in each case the Bekenstein-Hawking formula does not hold.  In section \ref{qessec} we review the QES formula and its use to compute a unitary Page curve in \cite{Almheiri:2019psf,Penington:2019npb}.  In section \ref{nosymsec} we show how a unitary Page curve calculation of this type forbids global symmetries, and we also comment on the extension of our argument to higher-form global symmetries and the related conjectures of the completeness of gauge representations and the compactness of internal gauge groups.  Finally in section \ref{finalsec} we discuss the general relationship of Euclidean quantum gravity to holography.

\section{Low-dimensional examples of global symmetry in gravity}\label{lowdimsec}
We begin by reviewing low-dimensional examples of quantum gravity theories which have global symmetries.  The first is the worldline theory of a free relativistic particle moving in $d$-dimensional Minkowski space \cite{Polchinski:1998rq}, with action 
\be
S=-m\int dt \sqrt{-\eta_{\mu\nu}\dot{X}^\mu\dot{X}^\nu}
\ee
for the embedding map $X^\mu(t)$.
To make this look more gravitational we can integrate in a $(0+1)$-dimensional dynamical metric $g=-e^2dt^2$, leading to
\begin{align}\nonumber
S&=\frac{1}{2}\int dt \left(e^{-1}\eta_{\mu\nu}\dot{X}^\mu \dot{X}^\nu-em^2\right)\\
&=-\frac{1}{2}\int dt \sqrt{-g_{tt}}\left(g^{tt} \eta_{\mu\nu}\dot{X}^\mu \dot{X}^\nu+m^2\right).
\end{align}
In the latter presentation we apparently have a $(0+1)$-dimensional quantum field theory coupled to dynamical gravity, but on the other hand this theory is trivially solvable and clearly possesses a large global symmetry consisting of the ``target space'' Poincar\'e transformations
\be
X^{\mu\,\prime}=\Lambda^\mu_{\phantom{\mu}\nu}X^\nu+a^\mu.
\ee 
There is no known sense in which this model has black holes, so they do not provide any obstruction to this global symmetry. We emphasize that the canonical quantization of this theory does not produce a sum over branching worldlines for the particle: these could be added by introducing new interactions, but doing so spoils the completeness of the theory and requires a non-perturbative completion into quantum field theory. 

A similar family of examples in $1+1$ dimensions is given by canonically quantizing the string worldsheet action 
\be
S=-\frac{1}{4\pi \alpha'}\int d^2x\sqrt{-g}g^{ab}\partial_a X^\mu \partial_b X_\mu,
\ee
which again has target space Poincar\'e transformations as a global symmetry and does not have black holes.  Canonical quantization again does not produce a sum over branching worldsheet topologies: one can be added by hand via explicit splitting/joining interactions, but doing so again renders the series divergent and requires a completion into some non-perturbative description of interacting string theory. 

A less trivial set of examples is constructed by coupling the CGHS/RST \cite{Callan:1992rs,Russo:1992ax} or Jackiw-Teitelboim \cite{Jackiw:1984je,Teitelboim:1983ux,Almheiri:2014cka} model of gravity in $1+1$ dimensions to any non-chiral ``matter'' conformal field theory with central charge $c=c_L=c_R$. These two cases are similar, so we choose to focus on the latter. JT gravity coupled to conformal matter has been studied in great detail in recent years: initially because of its close connection \cite{Jensen:2016pah,Maldacena:2016upp,Engelsoy:2016xyb,Kitaev:2017awl,Sarosi:2017ykf} to the SYK model \cite{Sachdev:1992fk,Kitaev,Polchinski:2016xgd,Maldacena:2016hyu}, and later as an interesting theory of gravity by itself \cite{Harlow:2018tqv,Lin:2018xkj,Yang:2018gdb,Saad:2019lba,Gross:2019ach, Gross:2019uxi, Iliesiu:2020zld, Stanford:2020qhm}.  The action is
\be
S=\int_M d^2x \sqrt{-g}\left(\Phi_0 R+\Phi(R+2)\right)+2\int_{\partial M}dt\sqrt{-\gamma}\left(\Phi_0 K+\Phi(K-1)\right)+S_{CFT}(\psi_i,g),
\ee
where $\Phi_0$ is a constant, $\Phi$ is a dynamical ``dilaton'' field,  $g_{\mu\nu}$ is a dynamical metric, $R$ is its Ricci scalar, $\psi_i$ are CFT matter fields that we emphasize do not couple to the dilaton $\Phi$, and $\gamma$ and $K$ are the induced metric and extrinsic curvature at the asymptotically-AdS boundary $\partial M$.  The boundary conditions at $\partial M$ are that 
\begin{align}\nonumber
\Phi|_{\partial M}&=r_c \phi_b\,,\\
\gamma|_{\partial M}&=-r_c^2dt^2,
\end{align}
with $r_c$ taken to infinity.

There are two methods which have been used to quantize JT gravity coupled to conformal matter.\footnote{In both quantization methods one takes the integration contour for the dilaton to include the range $ \Phi_0 + \Phi < 0$. This is perfectly well-defined from the two-dimensional perspective, but from a higher-dimensional perspective it  would imply integrating over negative metrics. It would be interesting to understand to what extent the results described below depend on this choice of contour.}  The first is canonical quantization, which from a path integral point of view is equivalent to summing over real globally-hyperbolic Lorentzian geometries.  Since this theory is renormalizable, canonical quantization leads to a well-defined quantum theory.  With two asymptotic boundaries and no matter the resulting theory is equivalent to the quantum mechanics of a particle moving in an exponential potential \cite{Harlow:2018tqv}, while with a nontrivial matter CFT it can be solved by Weyl transformation to flat space \cite{Almheiri:2019psf,Almheiri:2019yqk,Almheiri:2019qdq} (here ``solved'' means that all observables can be expressed in terms of the vacuum correlation functions of the matter CFT in flat space).  This quantization preserves any global symmetry of the matter CFT which does not have a mixed anomaly with diffeomorphism symmetry, and thus it gives another example of a quantum gravity theory with a global symmetry. For example we can take the matter CFT to be $N$ massless Dirac fermions, in which case there is a $U(N)$ global flavor symmetry.  Unlike the previous examples, this theory has black hole solutions.  We will see in the next section however that the entropy of these black holes is not compatible with the Bekenstein-Hawking formula, and their evaporation is not a unitary process in the usual sense (this theory has ``remnants'').

The other method for quantizing this theory takes the Euclidean gravity path integral as its starting point.  In a non-gravitational system this would be equivalent to canonical quantization, but, as emphasized in \cite{harlowkitp} and explained further in section \ref{finalsec} below, in a gravitational system they can be different.  In particular with no matter CFT, the Euclidean ``quantization'' of JT gravity (including a sum over genus) does not lead to a quantum mechanical system at all: instead it is an average over quantum systems \cite{Saad:2019lba,Stanford:2019vob,Johnson:2019eik,Johnson:2020heh}.   Including nontrivial conformal matter does not help: it introduces ``pinching'' divergences into the higher-genus contributions to the Euclidean partition function \cite{Saad:2019lba}, destroying the renormalizability which made the model well-defined in the first place.  On the other hand, there is strong evidence that in AdS/CFT topologically non-trivial Euclidean configurations which do not have a canonical interpretation must be included in the bulk description to correctly match the dual CFT \cite{Witten:1998qj,Jafferis:2017tiu,Almheiri:2019qdq,Penington:2019kki}.  Our current read of the situation is that in any theory of gravity which is rich enough to have black hole solutions, including topologically non-trivial Euclidean configurations in the gravitational path integral is consistent with quantum mechanics if and only if we are viewing that path integral as a low-energy effective theory which is UV-completed into a holographic system; we return to this point in section \ref{finalsec}.

Our final example of a quantum gravity theory with a global symmetry will be the oriented version of pure Einstein gravity in $2+1$ dimensions, which we will study with a negative cosmological constant:
\be
S=\frac{1}{16\pi G}\int_M d^3x \sqrt{-g}(R+2)+\frac{1}{8\pi G}\int_{\partial M}d^2x\sqrt{-\gamma}(K-1).
\ee 
This is also a very well-studied model, see e.g. \cite{Witten:1988hc,Banados:1992wn,Witten:2007kt,Maloney:2007ud,Maloney:2015ina,Kim:2015qoa,Maloney:2016gsg,Benjamin:2020mfz,Maxfield:2020ale}.  Its status based on a Euclidean starting point remains unclear (see e.g. \cite{Maloney:2007ud,Benjamin:2020mfz, Maxfield:2020ale}), but it can certainly be canonically quantized \cite{Witten:1988hc, Kim:2015qoa, Maloney:2015ina} and we will take that starting point here.  The point then is that in the Lorentzian path integral over globally-hyperbolic geometries consistent with the boundary conditions, we have a choice whether or not to include non-orientable spatial geometries.  This choice is equivalent to whether or not parity symmetry is gauged, so if we include only oriented geometries then parity is a global symmetry \cite{Harlow:2018tng}. (This is analogous to the role of worldsheet parity in string theory: for the oriented string worldsheet parity is a global symmetry, while for the unoriented string it is a gauge symmetry \cite{Polchinski:1998rq}.)  This theory again does not have unitary black hole evaporation: for one thing there are no local degrees of freedom for a black hole to evaporate into, and for another the number of microstates is not compatible with the Bekenstein-Hawking formula since the quantization of the moduli space at fixed genus leads to a continuous spectrum and the sum over spatial genus is also divergent \cite{Maloney:2015ina, Kim:2015qoa}.

\section{Quantum extremal surfaces and the black hole information problem}\label{qessec}
In the previous section we presented several examples of quantum gravity theories with global symmetry, each of which was based on a renormalizable path integral over metrics.  Unfortunately Einstein gravity is not renormalizable in $3+1$ dimensions (or in $2+1$ dimensions with matter fields), and so far the ``asymptotic safety'' program that looks for a strongly-coupled UV fixed point for Einstein gravity in $3+1$ dimensions (such as the proponents of loop quantum gravity hope to find) has been unsuccessful.  Moreover even if such a program were successful, the above examples suggest that it would lead to black holes whose entropy is not consistent with the Bekenstein-Hawking formula \eqref{BHform}, basically because locality would ensure the validity of UV/IR decoupling so one would be able to explicitly construct remnants.  Unlike the remnants we will soon meet in lower-dimensional gravity, which are manifest already in the low-energy variables since the relevant black holes have horizons even in the zero-energy limit, these remnants would necessarily involve high-energy degrees of freedom in some essential way.  

A somewhat more philosophical way to think about this is to note that in order for the black hole information problem to have a simple operational realization, we would like a theory in which 1) the natural final state of the evaporation is non-singular and horizon-free and thus can be expected to have zero entropy (except for any entropy in the Hawking radiation) and 2) there exist propagating degrees of freedom from which a black hole can be formed and into which it can evaporate.  JT gravity coupled to conformal matter and the CGHS/RST model both fail condition 1), while pure Einstein gravity in $2+1$ dimensions fails condition 2).  We are unaware of any renormalizable theory of gravity which satisfies both conditions, and we thus do not view the existence of these low-dimensional theories as evidence for the existence of a viable asymptotic safety scenario which resolves the $3+1$ dimensional information problem via remnants.  Indeed so far all examples of UV-complete quantum gravity satisfying these conditions have instead come from string theory \cite{Polchinski:1998rq}, and those which are well-defined non-perturbatively are \textit{holographic} in nature \cite{tHooft:1993dmi,Susskind:1994vu,Banks:1996vh,Maldacena:1997re, Itzhaki:1998dd}: their fundamental description lives at some asymptotic boundary in a lower number of dimensions, and in particular is not based on a local Lagrangian living in the gravitational spacetime that emerges at low energies.  And indeed in these examples there are no remnants: the Bekenstein-Hawking formula \eqref{BHform} holds for all black holes for which it has been checked, see e.g. \cite{Strominger:1996sh,Strominger:1997eq,Benini:2015eyy, Shaghoulian:2015lcn}. In such theories the gravitational path integral should be regarded as a low-energy effective description, which can capture some features of what is going on but not all of them.  

The best-understood examples of holographic theories of quantum gravity are those provided by the AdS/CFT correspondence.  One of the great successes of that correspondence has been the development over the last 15 years, beginning with the seminal work of Ryu and Takayanagi \cite{Ryu:2006bv}, of a bulk formula for competing the microscopic von Neumann entropy of a boundary subregion.  The final version of this formula is called the quantum extremal surface (QES) formula \cite{Engelhardt:2014gca}, and it says that for any boundary CFT spatial subregion $R$ and any reasonably semiclassical state $\rho$ in the CFT Hilbert space the von Neumann entropy of $\rho$ on $R$ is given (for Einstein gravity coupled to matter) by
\be\label{QES}
S(\rho_R)=\min_{\gamma_{R}}\underset{\gamma_R}{\mathrm{ext}}\left(\frac{\mathrm{Area}(\gamma_R)}{4G}+S(\rho_{\gamma_R})\right).
\ee
Here the extremum (and the minimum if the extremum is not unique) are taken over codimension two surfaces $\gamma_R$ which are anchored to $R$ in the sense that $\partial \gamma_R=\partial R$, and which are also homologous to $R$ in the sense that there is a spatial surface $H_{\gamma_R}$ such that $\partial H_{\gamma_R}=\gamma_R\cup R$, and $\rho_{\gamma_R}$ is the state of the bulk fields on $H_{\gamma_R}$ in the low-energy effective description.  The minimal extremal surface $\gamma_R$ is called the \textit{quantum extremal surface}, and the bulk domain of dependence of $H_{\gamma_R}$,
\be
W_R\equiv D[H_{\gamma_R}],
\ee
is called the \textit{entanglement wedge of $R$}.  General information-theoretic arguments show that in any situation where the QES formula holds, all bulk operators in $W_R$ (more carefully the intersection of the entanglement wedges over all states in the code subspace) can be represented in the dual CFT as operators with support only on $R$ \cite{Dong:2016eik,Cotler:2017erl, Hayden:2018khn}. This statement is called \textit{entanglement wedge reconstruction} \cite{Czech:2012bh,Wall:2012uf,Headrick:2014cta}.  

In recent years, it has gradually been understood that the QES formula can be used for more than just computing the von Neumann entropy of a holographic CFT subregion $R$.  In \cite{Lewkowycz:2013nqa}, it was already noted that the Euclidean replica method gives a fairly general bulk algorithm for deriving entropy formulas that do not seem to rely on the details of AdS/CFT.  It was also noticed in \cite{Swingle:2009bg,Pastawski:2015qua,Hayden:2016cfa} that a version of the QES formula holds in rather general tensor network constructions, and it was then observed that this extends to a version which holds in essentially all finite-dimensional quantum error correcting codes \cite{Harlow:2016vwg} (aspects of this result were extended to an approximate setting in \cite{Cotler:2017erl,Hayden:2018khn} and to infinite dimensions in \cite{Kang:2018xqy,Faulkner:2020hzi}).  It was noted in \cite{Hayden:2018khn} that this general QES formula can be applied to an arbitrary reservoir system which is entangled with a holographic CFT, provided that we allow the ``entanglement wedge'' of the reservoir to include regions which are in the gravitational part of the system.  Then in \cite{Penington:2019npb} (and more implicitly in \cite{Almheiri:2019psf})  it was shown that once such regions are accounted for, which in \cite{Almheiri:2019hni} were later christened ``islands,'' the QES formula can be used to demonstrate a unitary Page curve for an evaporating black hole.  Finally in  \cite{Almheiri:2019qdq,Penington:2019kki} it was pointed out that a careful application of the Euclidean replica argument of \cite{Lewkowycz:2013nqa} leads to ``replica wormholes" -- topologically nontrivial gravitational configurations connecting the various sheets of the replica -- which imply the appearance of islands in the QES formula when it is applied to a non-gravitational reservoir interacting with a gravitational system. This argument assumes  it is correct to include higher Euclidean topologies in the path integral.  As mentioned above and explained further in section \ref{finalsec}, we expect that this inclusion is justified for any gravitational effective field theory which is UV-completed into a holographic theory; this therefore justifies applying the QES formula to the reservoir in such theories as well.

\bfig
\includegraphics[height=7cm]{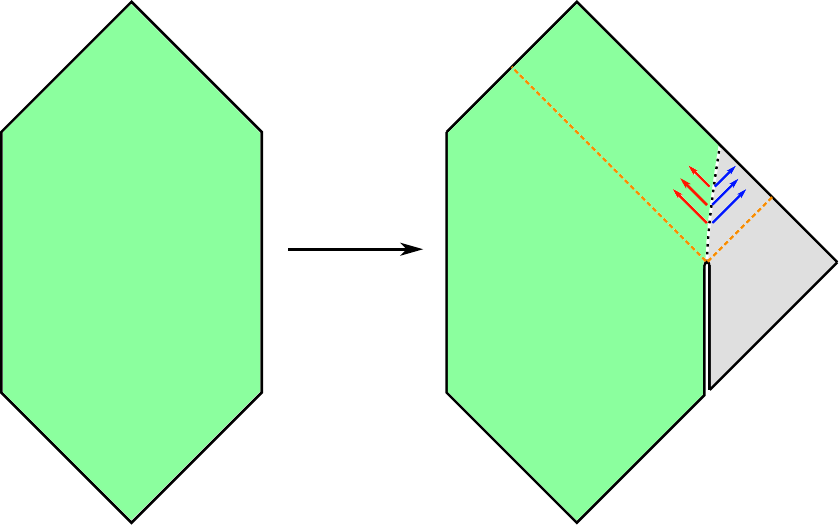}
\caption{The evaporating wormhole in JT gravity studied in \cite{Almheiri:2019psf}.  Turning on the interaction between the green gravitational region and the grey reservoir region produces two positive energy shells, shown as orange dashed lines.  After this,  positive energy Hawking radiation (shown in blue) leaks out into the reservoir system, while negative energy radiation (shown in red) falls into the right black hole and gradually decreases its energy.}\label{2devapfig}
\efig
This general version of the QES formula is what we will use in the following section to forbid global symmetry in quantum gravity, so we will first quickly review its application to an evaporating black hole in JT gravity coupled to conformal matter as in \cite{Almheiri:2019psf}.  The standard two-sided wormhole solution of pure JT gravity in Schwarzschild coordinates is given by
\begin{align}\nonumber
\Phi&=\phi_b r,\\
ds^2&=-(r^2-4\pi^2T^2)dt^2+\frac{dr^2}{r^2-4\pi^2T^2}.\label{JTbh}
\end{align}
Here $T$ is the Hawking temperature, the horizon is at $r_s=2\pi T$,
and the analogue of the Bekenstein-Hawking entropy of one side is
\be\label{SJT}
S=4\pi (\Phi_0+\Phi(r_s))=4\pi (\Phi_0+2\pi \phi_b T).
\ee
The global geometry is shown on the left side of figure \ref{2devapfig}; the future and past null boundaries are Cauchy horizons indicating the maximal extent of the causal development which is determined by the initial state and the choice of boundary conditions.  Adding conformal matter to this solution does not do much; in the thermofield-double state the matter state on each side is thermal and gives an $O(c)$ correction to the entropy \eqref{SJT}.  In \cite{Almheiri:2019psf} it was observed that the situation becomes much more interesting if at $t=0$ we couple one of the two boundaries to a reservoir system consisting of the same matter CFT propagating (without the metric or dilaton) on a half-space and starting in the vacuum.  This allows energy to gradually leak out into the reservoir system, decreasing the energy of the remaining black hole on that side of the wormhole (see the right side of figure \ref{2devapfig} for an illustration).   

One of the main insights from \cite{Almheiri:2019psf} is that there are two different ways of computing the von Neumann entropy of the reservoir system as a function of time.  The first approach is to view this renormalizable bulk theory as a complete theory of quantum gravity which can be solved by Weyl transformation, as we did in the previous section where we noted that the matter CFT may have global symmetries.  This leads to a time dependence
\be\label{Sres}
S_{\text{res}}(t)=16\pi^2\phi_bT_1\left(1-e^{-\frac{ct}{96\pi\phi_b}}\right)
\ee 
for the von Neumann entropy of the reservoir, where $T_1$ is the temperature of the right black hole after the orange shell in figure \ref{2devapfig} falls in.  This however leads to a version of the black hole information problem: at late times the black hole on the right side of the wormhole has evaporated down to zero temperature, so from \eqref{SJT} the coarse-grained entropy of the two black holes together is
\be\label{Scoarse}
S_{\text{coarse}}=8\pi \Phi_0+8\pi^2\phi_bT_0,
\ee
where $T_0$ is the initial temperature of the two sides prior to turning on the coupling to the reservoir. The key point is that if $\phi_b T_0\gg \Phi_0$, then for sufficiently late times we have
\be\label{infoloss}
S_{\text{res}}\gg S_{\text{coarse}},
\ee
which is impossible if we think that the reservoir system is purified by the two remaining black holes. (An analogous conclusion was reached in the CGHS/RST model in \cite{Fiola:1994ir}.)  The only way \eqref{infoloss} can be consistent with unitarity is if \eqref{SJT} is incorrect.  In particular note that by adding a pure ingoing matter shell to the reservoir system prior to coupling the systems, we can arrange for $T_1$ to be arbitrarily large compared to $T_0$, and thus to avoid a contradiction with unitarity we need the entropy for fixed $T_0$ to be arbitrarily large.  And indeed it is: this renormalizable bulk theory can have an arbitrarily large number of low-energy excitations near black hole horizon. In the context of the black hole information problem objects of finite energy and infinite entropy are referred to as remnants, so in this theory the ``resolution'' of the information problem is that there are remnants that do not obey \eqref{SJT}.  Therefore from the point of view of this paper this theory is allowed to have global symmetries, and indeed it will if the matter CFT does.  

\bfig
\includegraphics[height=8.5cm]{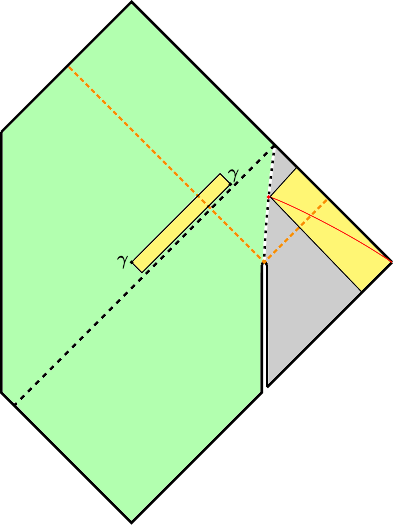}
\caption{The new quantum extremal surface $\gamma$ of \cite{Almheiri:2019psf,Penington:2019npb} for an evaporating JT wormhole.  The entanglement wedge of the reservoir at the boundary time indicated by the red dot is shaded yellow, note in particular the island in the gravitational region.}\label{2dislandfig}
\efig
The second approach to computing the entropy of $S_{\text{res}}$ is to instead view the bulk theory as a low-energy effective theory, to be UV-completed into some yet-to-be-determined holographic theory.  We may then use the QES formula \eqref{QES} to compute the entropy of the reservoir system (with $\frac{\mathrm{Area}}{4G}$ replaced by $4\pi (\Phi_0+\Phi)$).  At early times this calculation agrees with equation \eqref{Sres}, since then $S_{\text{res}}$ is small and there is no benefit in including any island contributions.  At late times however, \cite{Almheiri:2019psf,Penington:2019npb} made the remarkable discovery that there is a new candidate quantum extremal surface which has no classical counterpart.  The location of this surface is shown in figure \ref{2dislandfig}, with the entanglement wedge of the reservoir shaded yellow.  The value of the dilaton on this surface decreases with time as the black hole evaporates, and eventually it gives a candidate minimum in the QES formula which is smaller than the naive contribution without an island.  The formula \eqref{Sres} for the von Neumann entropy of the reservoir is thus modified to 
\be\label{page}
S_{\text{res}}(t)\approx \min\left[16\pi^2\phi_bT_1\left(1-e^{-\frac{ct}{96\pi\phi_b}}\right),8\pi \Phi_0+8\pi^2\phi_bT_0\left(1+\frac{T_1}{T_0}e^{-\frac{ct}{96\pi\phi_b}}\right)\right],
\ee
which differs from \eqref{Sres} precisely in those situations where there would have been an information problem, and in just such a way that at late times $S_{\text{res}}$ cannot exceed $S_{\text{coarse}}$!  Thus this result is consistent with the idea that the reservoir is entangled with the union of the remaining black holes, each having maximal entropy \eqref{SJT}.

The fact that the quantum extremal surface formula is able compute the unitary Page curve \eqref{page} using only semiclassical ingredients is a remarkable discovery.  On the other hand, since we are now only viewing the bulk description as a low-energy effective theory, we no longer know whether or not a global symmetry of the matter CFT is preserved in the UV completion.  We will now argue that in any situation where the Page curve can be derived in this manner, no global symmetry can exist in the full theory.

\section{A ``no global symmetries'' theorem from an evaporating black hole}\label{nosymsec}
\bfig
\includegraphics[height=5.5cm]{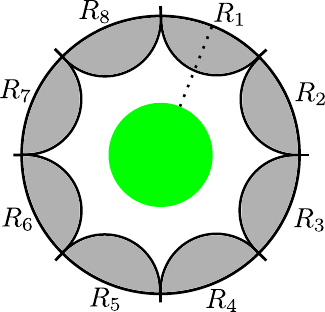} \hspace{1cm}\includegraphics[height=5.5cm]{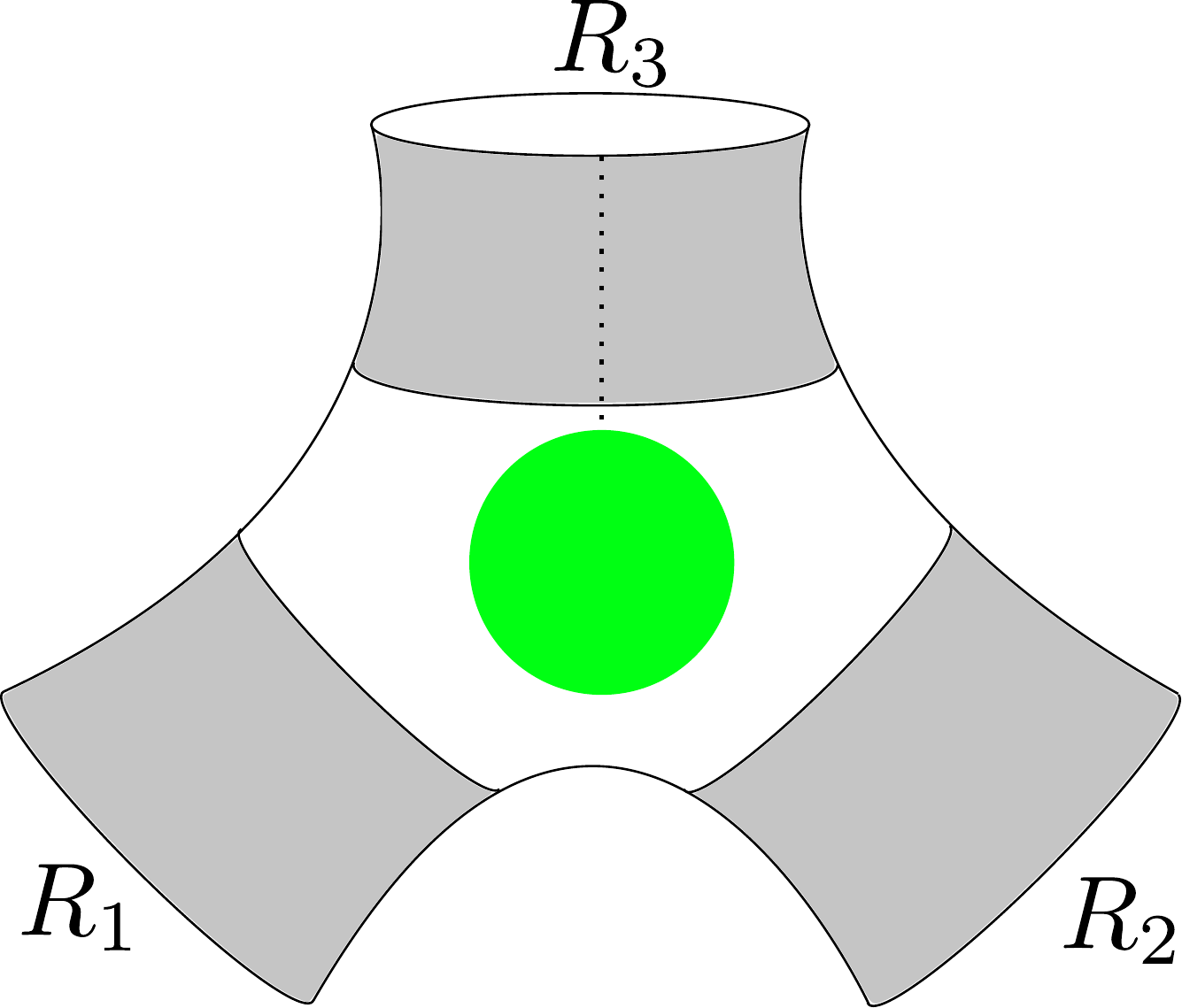}
\caption{Two situations where a bulk global symmetry would lead to a contradiction in AdS/CFT (figures from \cite{Harlow:2018tng}).  The green circle represents a putative bulk operator which is charged under the global symmetry, and the dotted line represents its (neutral) gravitational dressing.  The contradiction is that in the boundary CFT a global symmetry operator $U(g)$ must be a product of operators whose support is each at most one of the $R_i$, and each of those operators can therefore only act nontrivially on bulk operators in the entanglement wedge of that $R_i$.  Thus none of the operators, and therefore also their product, can act nontrivially on the green charged operator: it must be neutral.}\label{nosymfig}
\efig
In \cite{Harlow:2018jwu,Harlow:2018tng}, two versions of an argument were given that global symmetries are impossible in AdS/CFT.  The basic idea in both versions is to split the boundary into disjoint spatial regions $R_i$, such that there is a large region in the bulk which is not contained in the union $\cup_i W[R_i]$ of their entanglement wedges. This can be done either by splitting a single boundary into multiple regions or having multiple boundaries connected by a (spatial) wormhole; both options are shown in figure \ref{nosymfig}.  The key point is then that by the locality of the boundary CFT, any global symmetry group (restricting for simplicity to internal symmetries) must be represented on the CFT Hilbert space by a family of unitary operators $U(g)$ which have a product structure
\be
U(g)=\prod_i U(g,R_i) U_{edge},
\ee
where here $U(g,R_i)$ is a unitary operator which implements the symmetry $U(g)$ on operators in $R_i$ but does nothing in its spatial complement $\ol{R}_i$, and $U_{edge}$ is a unitary operator with support only on the boundaries of the $R_i$ that reflects the short-distance edge ambiguity inherent in $U(g,R_i)$ (in the wormhole situation there is no $U_{edge}$; see \cite{Harlow:2018tng} for much more discussion on why $U(g)$ must have this structure in quantum field theory).  The point is then that each $U(g,R_i)$, and also $U_{edge}$, has support only on a boundary region whose entanglement wedge does not contain a large region of the bulk (see figure \ref{nosymfig}).  On the other hand, if we had a nontrivial bulk global symmetry then there would need to be some bulk operator which is charged under it.  We could put that charged operator in the region which none of the $U(g,R_i)$ or $U_{edge}$ can access, and then it would be impossible for that operator to transform under conjugation by $U(g)$.  Thus there can be no such operator, and therefore no global symmetry in the first place.  

As presented, this argument relies on details of the AdS/CFT correspondence.  On the other hand, it feels more general - in any holographic theory where the microscopic description lives far away in a lower number of dimensions a similar contradiction seems like it should be possible. Our goal now is to present such a contradiction. Indeed let $S$ denote a quantum gravity system which is sufficiently semiclassical to allow for the existence of black holes that are large compared to the Planck scale.  Moreover let $R$ denote a ``reservoir'' system consisting of weakly-interacting quantum fields propagating on $\mathbb{R}^d$ (possibly including linearized gravitons), where $d$ is the spacetime dimension of the semiclassical description of $S$.  We now make three assumptions:
\bi
\item We can couple $R$ and $S$ together in such a way that an initial state consisting of a pure state black hole in $S$ and the vacuum in $R$  evolves unitarily. This evolution can be described semiclassically as the black hole producing Hawking radiation which is then gradually transferred into $R$. 
\item For at least one initial state of the black hole, e.g. one formed by a fast collapse, the fine-grained entropy of any subregion of $R$ at later times can be computed using the quantum extremal surface formula (including possible islands in $S$), and the fine-grained entropy of $S$ can also be computed this way.
\item The coupling between $R$ and $S$ preserves any internal global symmetries of $S$, and moreover the action of these global symmetries can be extended to $R$ in a way that respects the locality of $R$ (meaning it sends any operator on any subregion of $R$ to another operator on the same subregion).   
\ei
All three of these assumptions apply to the situation of a holographic CFT $S$ coupled to a reservoir $R$ as in the setups of \cite{Almheiri:2019psf,Penington:2019npb}, but we expect them to hold far more generally.  Roughly speaking we can just think of $S$ as the black hole and $R$ as its Hawking radiation, in which case we expect these assumptions to be true in any situation where a black hole formed by a fast collapse evaporates via a unitarized version of the Hawking process. (We are less sure about more generic initial states of the black hole, as currently we do not have a global semiclassical picture of the geometry in which to apply the QES formula.)

\bfig
\includegraphics[height=6cm]{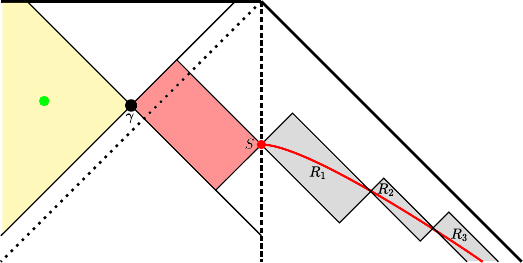}
\caption{The semiclassical picture of a situation where a global symmetry in a quantum gravity theory with unitary black hole evaporation leads to a contradiction. The solid red line indicates a Cauchy slice of the reservoir system $R$, the red dot indicates the quantum gravity system $S$, the black dot is the quantum extremal surface $\gamma$ found in \cite{Almheiri:2019psf,Penington:2019npb}, the vertical dashed line is the boundary between the gravitational and non-gravitational parts of the semiclassial description, and $R_1, R_2,\ldots$ are subregions of $R$.  The entanglement wedges of the $R_i$ are shaded grey, the entanglement wedge of $S$ is shaded pink, and the ``island'' which is part of the entanglement wedge of $R$ is shaded yellow.  The green dot represents an operator which is charged under the putative global symmetry.  Since the global symmetry operator $U(g)$ is a product of an operator on $S$ and operators on the $R_i$, none of its pieces can act nontrivially on the operator at the green dot and thus that operator must be neutral.}\label{evapsymfig}
\efig
Given these assumptions, we can now imitate the argument of \cite{Harlow:2018jwu,Harlow:2018tng} against global symmetry.  Namely we split up the reservoir system into pieces $R_i$ which are small enough that even at late times the entanglement wedges of the $R_i$ do not include any islands in the gravitational region (see figure \ref{evapsymfig}).  Since the global symmetry is preserved on the joint system and respects the locality of $R$, the unitary operators which represent it on the full system can be written as
\be\label{Urep}
U(g)=U(g,S)\prod_iU(g,R_i) U_{edge},
\ee
where $U(g,S)$ implements the symmetry on $S$ and does nothing on $R$, $U(g,R_i)$ implements the symmetry on $R_i$ and does nothing on $S$ or $R_j$ with $j\neq i$, and $U_{edge}$ has support only at the edges of the $R_i$.  But, as shown in figure \ref{evapsymfig}, at sufficiently late times there will be an island (shaded yellow) which is \textit{not} contained in the entanglement wedges of $S$ or any of the $R_i$ but \textit{is} contained in the entanglement wedge of $R$ \cite{Almheiri:2019psf,Penington:2019npb}.   If there were a global symmetry then we could put an operator which is charged under that global symmetry in this island.  This however would be inconsistent with the expression \eqref{Urep} for the symmetry operator $U(g)$: by entanglement wedge reconstruction the charged operator has to commute with each ingredient of $U(g)$, and thus with $U(g)$ itself, and therefore the operator must be neutral.  Thus no global symmetry could have existed in the first place.  

Both this argument and that of \cite{Harlow:2018jwu,Harlow:2018tng} rely on manifestations of the idea that entanglement can create geometry \cite{VanRaamsdonk:2010pw,Maldacena:2013xja}. Indeed if we use the model of \cite{Akers:2019nfi} to replace the pieces of the Hawking radiation with all but one of the exits of a multiboundary wormhole, with the black hole as the remaining exit, then our argument against global symmetries is the same as that in the right diagram of figure \ref{nosymfig}.  Quantum error correction is the precise mathematical formulation of the idea that entanglement can create geometry \cite{Almheiri:2014lwa}, and in the case of continuous symmetry groups both our argument and that of \cite{Harlow:2018jwu,Harlow:2018tng} are essentially recastings of the Eastin-Knill theorem forbidding the existence of continuous logical symmetries acting transitively on the physical degrees of freedom of a quantum error-correcting code \cite{eastin2009restrictions} (we emphasize however that our argument and that of \cite{Harlow:2018jwu,Harlow:2018tng} apply to discrete symmetries as well).  The importance of quantum error correction in interpreting the Page curve calculations of \cite{Almheiri:2019psf,Penington:2019npb} was emphasized in \cite{Penington:2019npb,Akers:2019nfi}, and to the extent that those calculations generalize to arbitrary holographic theories of quantum gravity, as we are assuming here, then we expect a role for quantum error correction in any such theory.

A natural question is whether or not this argument extends to excluding the ``generalized global symmetries'' of \cite{Gaiotto:2014kfa}.  We expect yes: to exclude a $p$-form global symmetry, one can apply our argument to the evaporation of a black hole with horizon topology $\mathbb{S}^{d-p-2}\times \mathbb{T}^{p}$, where $d$ is the spacetime dimension of the effective gravitational field theory.  For Einstein gravity in asymptotically-AdS space such solutions were constructed in appendix I of \cite{Harlow:2018tng}.  In any effective gravitational theory where such solutions exist, following \cite{Harlow:2018tng} we can dimensionally reduce the $\mathbb{T}^p$ and then apply our argument excluding ordinary ``zero-form'' global symmetries to the reduced theory to exclude $p$-form global symmetries as well.
 
In addition to the conjecture that there are no global symmetries, AdS/CFT arguments were also given in \cite{Harlow:2018jwu, Harlow:2018tng} for two related conjectures: the ``completeness hypothesis'' that in quantum gravity dynamical objects in all possible gauge representations must exist \cite{Misner:1957mt,Polchinski:2003bq}, and the ``compactness hypothesis'' that in quantum gravity all internal gauge groups must be compact \cite{Banks:2010zn}.  The AdS/CFT argument for the completeness hypothesis, based on an argument given in \cite{Harlow:2015lma} for the case of a $U(1)$ gauge group, uses only the minimal ingredients of the thermofield double state and the tensor-factorization of the microscopic theory with two asymptotic boundaries.  We expect that these ingredients should generalize directly to any holographic theory, so it does not seem necessary to consider black hole evaporation.\footnote{One can still try to replace the second boundary with a cloud of Hawking radiation, but one then meets the puzzling question of whether or not a Wilson line can enter an old black hole through its horizon and then come out directly into its Hawking radiation.  In the evaporation model of \cite{Akers:2019nfi} the answer to this question is ``yes,'' but for a real evaporating black hole we are less sure.}  The AdS/CFT argument for the compactness hypothesis makes more use of the local structure of the dual CFT, in particular the idea that the local operator algebra should be finitely generated. Extending this to more general theories of quantum gravity requires a further assumption along the lines that ``every black hole can be made from a finite number of ingredients,'' although we will not attempt to formalize this here.  It would be interesting to apply these ideas to other ``swampland'' proposals such as the weak gravity \cite{ArkaniHamed:2006dz} and distance \cite{Ooguri:2006in} conjectures; we leave this to future work.

\section{Euclidean gravity and holography}\label{finalsec}
The strongest conceptual motivation for holography as the ultimate description of quantum gravity comes from taking seriously the Bekenstein-Hawking formula \eqref{BHform} as dictating the size of the Hilbert space of the black hole \cite{tHooft:1993dmi, Susskind:1994vu}. So far this has had its sharpest incarnation in the context of string theory \cite{Banks:1996vh, Maldacena:1997re, Itzhaki:1998dd}, with the paradigmatic example being AdS/CFT. In that context, many questions can be sharply formulated and theorems -- like the absence of global symmetries in the bulk -- can be proven \cite{Harlow:2018jwu, Harlow:2018tng}. Given the picture that has recently emerged for unitarity of black hole evaporation, one is left to wonder whether we can strip away the baggage of AdS/CFT and return to the starting point: black holes, the Bekenstein-Hawking formula, and unitary evaporation. We have seen that to rule out the possibility of global symmetries in this context, we have had to make assumptions very similar to those motivating the idea of holography. Moreover we have seen that there are various ``quantum theories of gravity" which do have global symmetries, but they are not holographic and when they have black holes their entropy does not obey the Bekenstein-Hawking formula. 

A related point we have already mentioned is the role of the Euclidean path integral in gravitational effective field theory, which is the tool that gives \eqref{BHform} with the most generality. Through various examples we have tried to argue that the validity of this tool is very closely tied to whether the theory under consideration is holographic. We now promote this to a general conjecture:
\begin{quote}
The Euclidean path integral in a gravitational effective field theory with a quantum-mechanical UV completion correctly
computes von Neumann entropies such as (1.1) if and only if that UV completion is holographic, in which
case the entropies are those of the holographic theory.
\end{quote}
As motivation for this conjecture, we first recall that in ordinary quantum field theory on a spatial manifold $\Sigma$ the Euclidean path integral representation of a thermal partition function is derived by inserting complete sets of states into a thermal trace
\be
Z=\Tr\left(e^{-\beta H}\right),
\ee
which leads to a path integral on the manifold $\mathbb{S}^1\times \Sigma$.  Applying this algorithm to a renormalizable gravitational field theory such as Jackiw-Teitelboim gravity coupled to conformal matter therefore only includes manifolds which are topologically of the form $\mathbb{S}^1\times \Sigma$ for some $\Sigma$.  As originally explained by Hawking, by time-translation symmetry the on-shell action of any gravitational field theory on such a manifold will be proportional  to $\beta$, giving a vanishing thermal entropy
\be
S(\beta)=(1-\beta\partial_\beta)\log Z=0
\ee
at leading order in the gravitational constant.  Therefore no time-translation-preserving saddle-point approximation to a Euclidean path integral derived from canonical quantization of a gravitational field theory can ever lead to the Bekenstein-Hawking formula \eqref{BHform} (or its generalizations from the Wald formula such as \eqref{SJT}).  There is however a standard proposal for how to fix this: instead of just including topologies of the form $M=\mathbb{S}^1\times \Sigma$, include all topologies $M$ such that $\partial M=\mathbb{S}^1\times \partial \Sigma$, where $\partial \Sigma$ is the topology of the spatial boundary \cite{Gibbons:1976ue}, even though most of these topologies are not generated by canonical quantization of the gravity variables.  In particular the boundary circle $\mathbb{S}^1$ is allowed to contract somewhere inside $M$, which invalidates Hawking's argument that $\log Z\propto \beta$ (his argument still shows that the entropy will be determined entirely by the fields in the vicinity of the surface where the circle contracts). The Euclidean Schwarzschild geometry has such a point where the thermal circle contracts to zero size, and evaluating its action leads directly to \eqref{BHform} (the Euclidean version of the JT black hole \eqref{JTbh} similarly leads directly to \eqref{SJT}).  We illustrate geometries of both types in figure \ref{eucfig}.  

\bfig
\includegraphics[height=3.5cm]{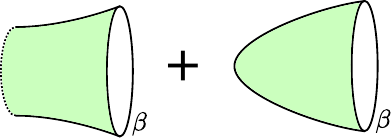}
\caption{Two contributions to the Euclidean gravity path integral with a boundary thermal circle of circumference $\beta$.  On the left some cycle of the transverse directions contracts at the dotted line, while on the right it is the thermal circle which contracts.  The contribution on the left is what is obtained from canonical quantization of gravitational effective field theory, and gives no contribution to the entropy at leading order in the gravitational coupling.  The contribution on the right leads to the black hole entropy \eqref{BHform}; it should be included only in effective theories which are UV-completed into a holographic description.}\label{eucfig}
\efig
Why though are we allowed to include geometries where the circle contracts?  We believe that the reason is holography: if the true microscopic description lives at the asymptotic boundary, then so does the true thermal circle!  Therefore we should really only require a product spacetime topology $\mathbb{S}^1\times \partial\Sigma$  at the boundary, and it is thus plausible to include geometries where the boundary thermal circle contracts in the interior of the spacetime.  In fact in AdS/CFT it is necessary to include them, as one is otherwise unable to obtain the correct scaling of the high-temperature entropy with the number of boundary degrees of freedom \cite{Witten:1998qj}.  In the models of holography based on stacks of Dp-branes this topological distinction and its implication for the entropy is captured in the boundary theory by the pattern of higher-form symmetry breaking \cite{Shaghoulian:2016xbx, Shaghoulian:2020omr}.  Another way to think about this is that although with these rules we include contributions which seem to violate the relationship between the Euclidean path integral and canonical quantization, that relationship only really needs to be preserved in the microscopic description at the boundary (and indeed in AdS/CFT the Lorentzian and Euclidean quantizations of the boundary CFT are equivalent).\footnote{This equivalence of Lorentzian and Euclidean quantizations only in the microscopic boundary theory is somewhat analogous to the factorization of multi-boundary thermal partition functions: this factorization must be true in any holographic theory, but it is not apparent in gravitational effective field theory even if we include arbitrary Euclidean topologies \cite{Giddings:1988cx,Maldacena:2004rf,ArkaniHamed:2007js,Harlow:2018tqv,Saad:2019lba,Penington:2019kki}.}  On the other hand, in any theory where we view the gravitational field theory description as fundamental --  as happens in our lower-dimensional examples of quantum gravity theories with global symmetries and we expect would happen in any asymptotic safety scenario -- then results (such as the Bekenstein-Hawking formula \eqref{BHform} and the QES formula \eqref{QES}) which rely on Euclidean topologies with no canonical interpretation need not be correct (and indeed they aren't in our examples). 

In holographic theories the successes of Euclidean quantum gravity once we allow arbitrary topologies to be included are undeniable, but they are also mysterious. How can the low-energy path integral know about the microstates of a black hole, which are expected to probe the deep ultraviolet? We do not have a complete answer to this question, which after all would likely require a non-perturbative understanding of the gravitational variables (for example a non-perturbative definition of string theory).  At least for thermodynamically-stable black holes in AdS/CFT however we are able to say something more: from the boundary point of view there is a high-temperature/low-temperature duality which relates geometries where the thermal circle contracts to geometries where it doesn't.  This is most familiar in the context of AdS$_3$, where it reduces to modular invariance in the boundary theory \cite{Strominger:1997eq}, but a similar picture is true in higher dimensions and for non-conformal models coming from stacks of Dp-branes with $p\neq 3$ \cite{Shaghoulian:2015lcn}.\footnote{These examples have some subtleties, for example for conformal theories in higher dimensions the theory at high temperature on a particular spatial manifold is generically related to the theory at low temperature on a different spatial manifold \cite{Cappelli:1988vw, Dolan:1998qk, Banerjee:2012gh, Shaghoulian:2015kta, Shaghoulian:2015lcn, Belin:2016yll, Shaghoulian:2016gol, Horowitz:2017ifu}.}  Since we do expect low-energy effective field theory to know the partition function on spacetimes where the thermal circle doesn't contract, the duality between low and high temperature ensures a reliable calculation of the high-temperature density of states within low-energy effective field theory.  One way to think about the magic of Euclidean gravity is thus that by allowing topologies where the thermal circle contracts, we manifestly restore a duality between low and high temperature which was not apparent from the canonical point of view.  What we lose however is a manifest interpretation of the gravity partition function as a thermal trace: so far we only know how to have manifest duality together with a manifest trace representation in the microscopic boundary description.

\paragraph{Acknowledgments} We thank Netta Engelhardt, Thomas Hartman, Daniel Jafferis, Hong Liu, Raghu Mahajan, and Wati Taylor for useful discussions.  This paper is dedicated to Leonard Susskind on the occasion of his 80th birthday: many of the ideas discussed here are his.  DH is supported by US Department of Energy Grant DE-SC0019127, the Simons Foundation as a member of the ``It from Qubit'' collaboration, the Sloan Foundation as a Sloan Fellow, the Packard Foundation as a Packard Fellow, and the Air Force Office of Scientific Research under the award number FA9550-19-1-0360. ES is supported in part by the Simons Foundation through the It From Qubit Collaboration (Grant
No. 38559) and QuantISED DE-SC0020360. 

\bibliographystyle{jhep}
\bibliography{bibliography}
\end{document}